# "Making you happy makes me happy"

# Measuring Individual Mood with Smartwatches

*Pascal Budner, Joscha Eirich, Peter A. Gloor*

*MIT Center for Collective Intelligence*

*pgloor@mit.edu*

**Abstract**

We introduce a system to measure individual happiness based on interpreting body sensors on smartwatches. In our prototype system we use a Pebble smartwatch to track activity, heartrate, light level, and GPS coordinates, and extend it with external information such as weather data, humidity, and day of the week. Training our machine learning-based mood prediction system using random forests with data manually entered into the smartwatch, we achieve prediction accuracy of up to 94%. We find that besides body signals, the weather data exerts a strong influence on mood. In addition our system also allows us to identify friends who are indicators of our positive or negative mood.

## 1. Introduction

When asked in a public Reddit session about his idea of success, Bill Gates responded[1] "Warren Buffett has always said the measure is whether the people close to you are happy and love you."

What if you could measure how happy you, and the people closest to you, are? In this paper we present a system that tracks individual mood of the wearer of a smartwatch, predicting mood through the sensors of the smartwatch and the associated smartphone.

Defining and measuring happiness has been an elusive goal for millennia. Aristotle already said "happiness is a state of activity". He saw it as a central purpose of human life and its pursuit as a goal in itself. Aristotle's happiness depends on the cultivation of virtue, although he was less strict in his definition of virtue than Confucius where virtue is fulfilling duties towards the state and the family. To achieve happiness according to Aristotle, a man needs to display the virtues of courage, generosity, justice, friendship, and citizenship (Aristotle 2004). As the highest of these virtues Aristotle considers "friendship of the good", i.e. friendship between those who resemble each other in virtue, and who love each other for themselves and not accidentally.

---

[1] https://www.reddit.com/r/IAmA/comments/5whpqs/im_bill_gates_cochair_of_the_bill_melinda_gates/

Science has come a long way since Aristotle, although his definition of happiness and friendship has not fundamentally changed over the last 2500 years. Only very recently have different scientists started to measure "honest signals" between individuals as indicators of happiness and friendship (Pentland & Heilbeck, 2010). Interaction between people has been measured using sociometric badges equipped with infrared and Bluetooth sensors, microphones, and accelerometers. To measure interaction between autistic people, a team at the MIT computer science lab used the sophisticated speech processing capabilities of the Samsung smartwatch. In particular, they built a wearable app that can parse conversation to identify the emotion in speech, combining it with tracking additional physical changes, such as increased skin temperature, heart rate, or movements (AlHanai & Ghassemi 2017). Other researchers have been using smartphones to track mood of their owners over extended periods of time and correlated it with their location (Doherty et al., 2014; Sandstrom et al. 2016). There are some potential limitations, though, in using exclusively the smartphone-embedded sensors to measure mood changes, for instance, reported mood states might be inaccurate, since people might not remember how they felt exactly during a certain time by reporting their mood once a day or once a week. Another group of researchers has used machine learning to correlate smartphone usage patterns with self-reported mood (Likamwa et al. 2013).

## 2. Happimeter: a Smartwatch-based Body Sensing System

In our own research we combine constantly recorded sensor data from a smartwatch with exogenous variables such as weather forecast and location. We developed our prototype system with the Pebble smartwatch. New users start with a generic machine-learning model, showing them the predicted mood (figure 1). If they disagree with the predicted mood, they can enter their correct mood directly on the smartwatch.

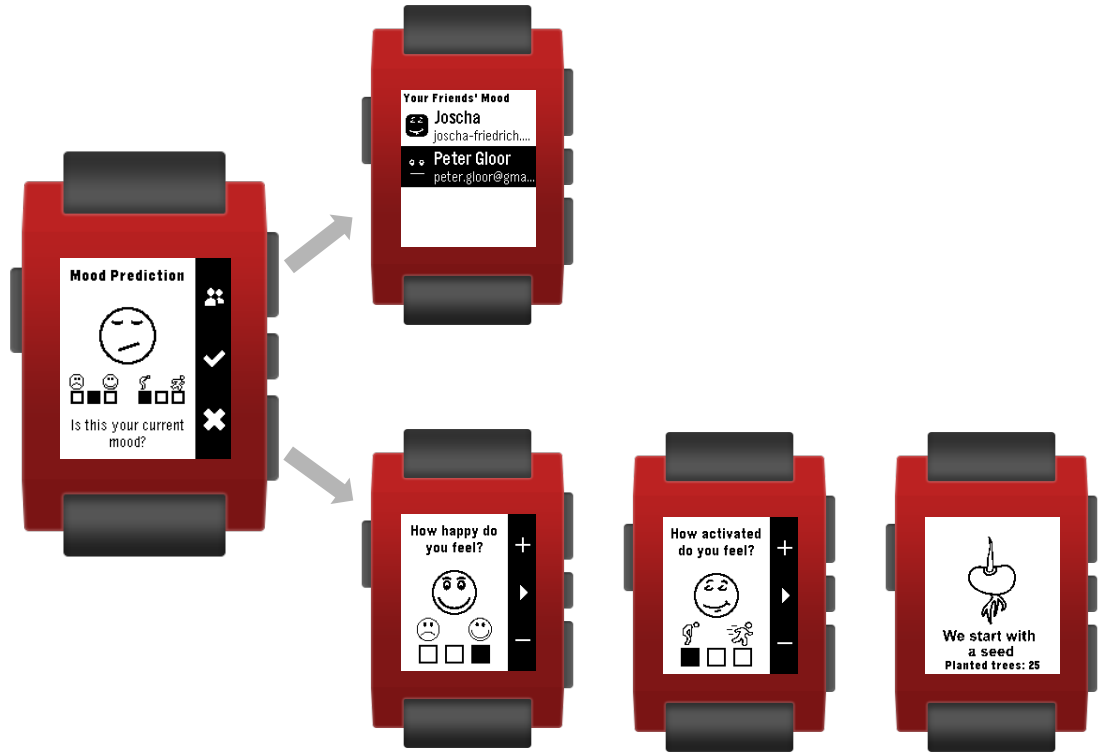

*Figure 1 Mood training process on Pebble smartwatch*

At the same time, users can also look at their friends' mood directly on the smartwatch (top of figure 1).

Besides the smartwatch app, we also built a smartphone app, which is available on the Google Play store, to visualize the user's data. Figure 2 illustrates five modules of the app. 2a allows the user to make mood inputs on the phone instead of the smartwatch. 2b shows how mood can be associated with the smartwatch-wearer's geolocation. 2c visualizes the social network of the user's friends that also use the Happimeter and are willing to share their mood with him. 2d shows statistics of the user's input data. 2e lists the friends of the user and their current mood. It is also possible to search for other Happimeter users, accept friend requests, control privacy options for mood-sharing, and unfriend users.

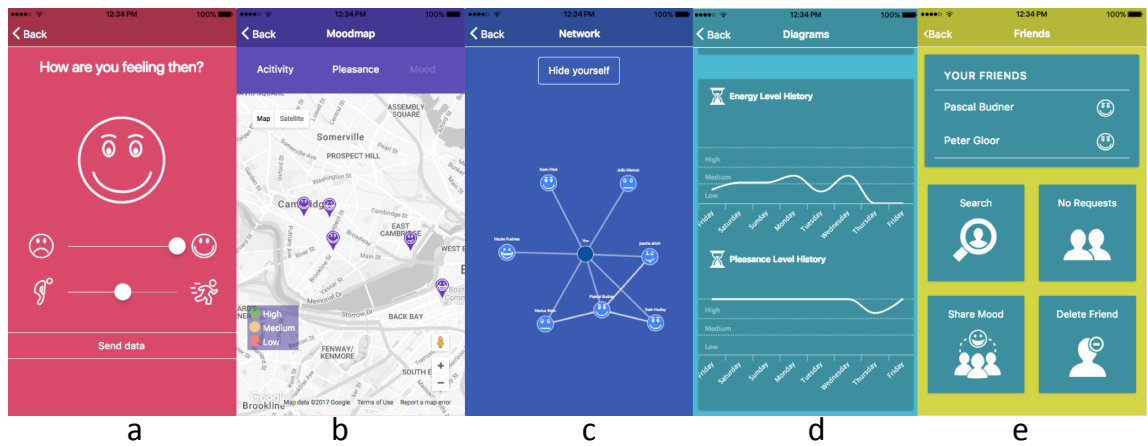

*Figure 2.* Modules of the smartwatch app presenting user feedback

Furthermore, a responsive website was built to allow users using other platforms the opportunity to review their data. The website consists of the same modules as the smartphone app. Figure 3 illustrates the dashboard providing immediate user feedback from the mood poll and the sensor readings. Based on the mood values provided by the users though the app, it reports the fluctuation in activation and pleasance, as well as heart rates in beats per minute (BPM) and an estimate of the user's personality traits if he/she took a personality test based on the International Personality Item Pool (Goldberg et al., 2006; Johnson, 2014).

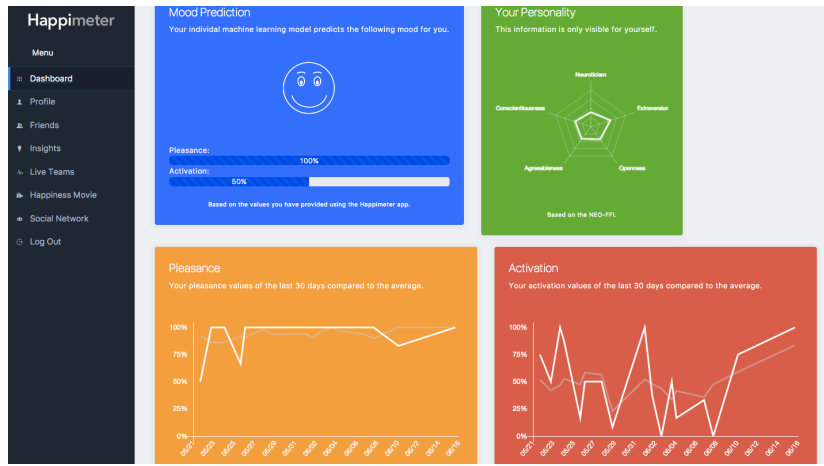

## 3. How to Measure Mood

We implemented a nine-outcome grid in the two dimensions "pleasance" and "activation" relying on the Circumplex Model of affect (Posner, Russell, & Peterson, 2005). As commonly done in social psychology, we therefore separate "happiness" into the arousal and valence dimensions. High pleasance means "happy", while highly activated means more aroused, as indicated by higher heart rate and blood pressure, being more sensory alert, and more ready to respond.

In the circumplex model the valence dimension (pleasant vs unpleasant) is on the horizontal axis and the activation dimension (active vs non-active) is on the vertical axis (Posner et al., 2005; Russell, 1980). Based on the three dimensions, where 2 represents high pleasance/activation, 1 medium pleasance/activation and 0 low pleasance/activation, we built a system that would ask two questions: (1) "How pleasant do you feel?" and (2) "How active do you feel?; The user chooses his pleasance and activation level with a scale from 1-3 and get immediate feedback in form of a single mood state. Examples of emotional states were: feeling happy, angry, relaxed, tired.

Similarly to the circumplex model of affect, we assume that all emotional states can be understood as a linear combination of two dimensions, one related to valence and the other to arousal or alertness (Barrett, 2006; Rafaeli, Rogers, & Revelle, 2007). Our model, described in Figure 3, reflects the assumption that specific emotions are connected to patterns of pleasance and activation within these two continua. As proposed by (Posner et al., 2005) joy could be conceptualized "*as an emotional state that is the product of strong Activation in the neural systems associated with positive valence or pleasure together with moderate Activation in the neural systems associated with arousal. Affective states other than joy likewise arise from the same two neurophysiological systems but differ in the degree or extent of Activation*".

We collect nine different mood states as a combination of three levels of pleasance and three levels of activation. This is similar to the approach followed by (Likamwa, Liu, Lane, & Zhong, 2013) in building MoodScope, a sensor that measures the mental state of the user based on how the smartphone is used.

On our nine-outcome grid as illustrated in Figure 4, we positioned happiness on an angle very close to high pleasance, based on the results of Russell (Russell, 1980) and Posner et al (Posner et al., 2005). As demonstrated by Russell (Russell, 1980), affective space is bipolar and antonyms are positioned approximately 180°: "*Beginning with happy at 7.8°, we can see that increases in angle at this point in the circle correspond to the increases in arousal and slight decreases in pleasure*" (Russell, 1980).

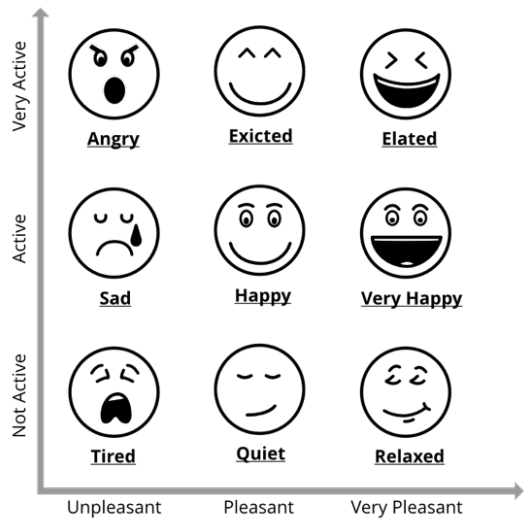

*Figure 4.* Nine-outcome grid used to elicit responses on mood states

In order to test the accuracy of the machine learning algorithms, which we use to assess simultaneously the dimensions of pleasure and activation, we introduce another variable called mood state. Mood state is a categorical variable created to classify the nine possible combinations of pleasure and activation. Mood state has the value of 1 if both pleasure and activation are 2; it has the value of 9 if both the dependent variables score 0. Each mood state can be described by a unique word. For instance, a pleasure value of 0 and an activation value of 2 represents the mood state 'angry' (coded as 3).

All variables are directly recorded by either the smartwatch or the weather API. The resulting feature set can be divided into two distinct groups: (1) body sensor data and (2) external influences (figure 5).

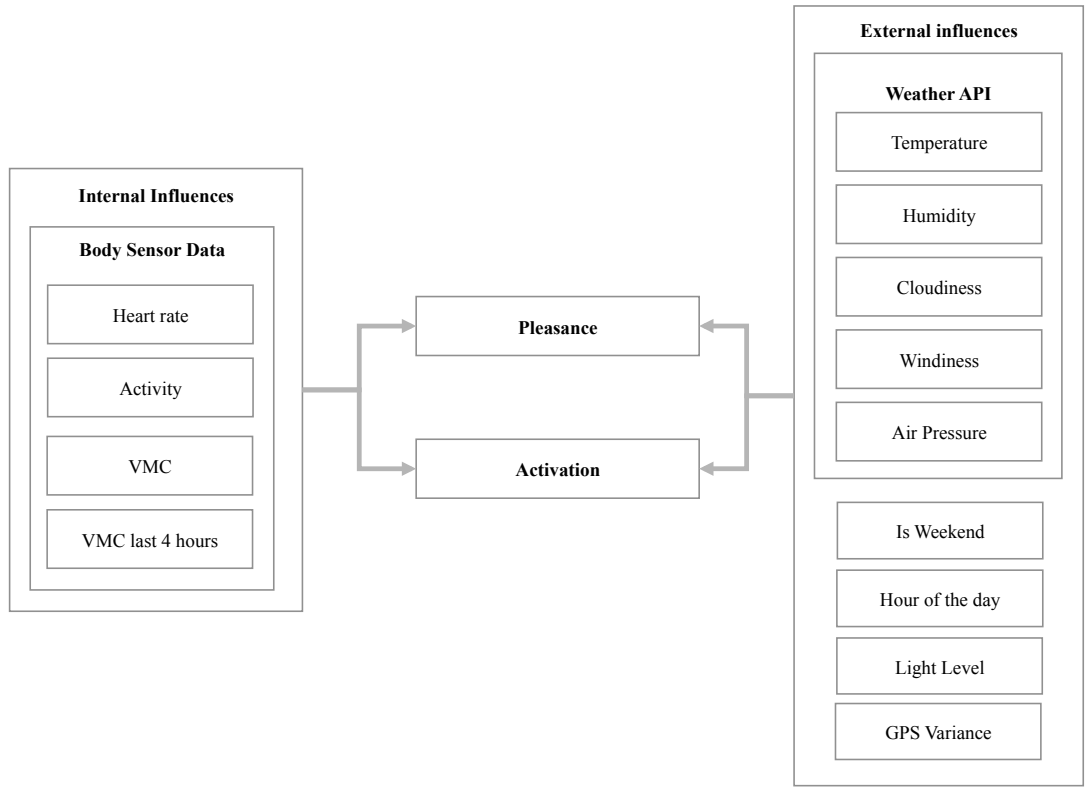

*Figure 5. Parameters used in the prediction model*

*(1) Body sensor data:*

This group contains sensor data that is only accessible by sensing the body of the smartwatch wearer. Heartrate measures the average number of heart beats per minute. Activity is a proprietary measure provided by the Pebble smartwatch and measures the activity level of the user. It ranges from low (0) to high (5) activity. VMC (vector magnitude counts) is a measure of the total amount of movement recorded by the smartwatch: more vigorous movement yields higher VMC values. "Activity in the last 4 hours" shows the average activity, which is indicated by the VMC, of the user during the last 4 hours.

*(2) External influences:*

External influences are features that are not provided by the body of the user but external circumstances such as time, position, or weather. "Is Weekend" indicates if data is collected on a week day or during the weekend. "Hour of the day" represents the local hour during which the data sample was collected. The light level measures the ambient light level at the moment of measurement and ranges from 0 (low) to 5 (high). "GPS variance" shows the change of the geological position of the user. Hence, a higher "GPS variance" implies visiting more distinct places. Furthermore, several weather measures are collected: temperature, humidity, cloudiness, windiness, and air pressure.

## 4. Verification of Mood Prediction Accuracy

To verify the accuracy of our system we recruited 60 people wearing the smartwatch from May 1st to June 30th, 2017. Our sample included graduate students, researchers, faculty members, consultants, and business industry leaders, with age ranging from 22 to 59. Their nationalities were German, Swiss, American, Chinese and Polish. Table 1 shows that 14 (42.42%) were female and 19 (57.58%) male. Almost 50% of all participants are under 30 years. When downloading the Happimeter app, participants agreed to participate in this study and received instructions via email. Participants were polled four times per day at a random time spread out through their awake hours by a vibration of the smartwatch, and were prompted to enter their mood states.

| Gender | | | | |
|---|---|---|---|---|
| **Female** | | **Male** | | **Unknown** |
| 14 (23.33%) | | 19 (31.67%) | | 27 (45%) |
| **Age** | | | | |
| **20 – 29** | **30 – 39** | **40 - 49** | **>50** | **Unknown** |
| 16 (26.67%) | 11 (18.33%) | 3 (5%) | 3 (5%) | 27 (45%) |

*Table 1. Distribution, Mean- and SD- Values of the participants, N = 60*

We are aware of the biases of a voluntary response sample compared to a random sample, as some members of the intended population are less likely to be included than others. At the same time, our goal was not to make inference on how the body sensors would affect those people. Our study aims at exploring a new methodology to recognize mood changes based on data recorded through smartwatches.

| | **High** | **Medium** | **Low** | **Total** | **Mean** | **Standard Deviation** |
|---|---|---|---|---|---|---|
| **Pleasance** | 13,436 (79.94%) | 2,855 (16.98%) | 515 (3.08%) | 16,806 (100%) | 1.7688 | 0.4889 |
| **Activation** | 2,662 (15.84%) | 9,784 (58.22%) | 4,360 (25.94%) | | 0.8989 | 0.6384 |

*Table 2. Descriptive statistics of dependent variables, N = 16,806*

Within three months, sensor data was collected and sent to the server every 15 minutes. Participants were encouraged to make mood inputs four times a day at random times. In addition, users were also able to make mood inputs manually whenever they wanted. All in all, the collected data sums up to 16,806 observations (a combination of mood input and sensor data entry) in total. During the whole experiment, almost 80% of all mood inputs show that the participants felt very pleasant (table 2). At the same time only 15.84% felt very active. Only 3.08% of the participants felt unpleasant. 25.94 % of the time they felt not active. Considering the variable pleasance, the experimental group was skewed toward high pleasance.[2] The pattern of the activation inputs indicates a normal distribution.

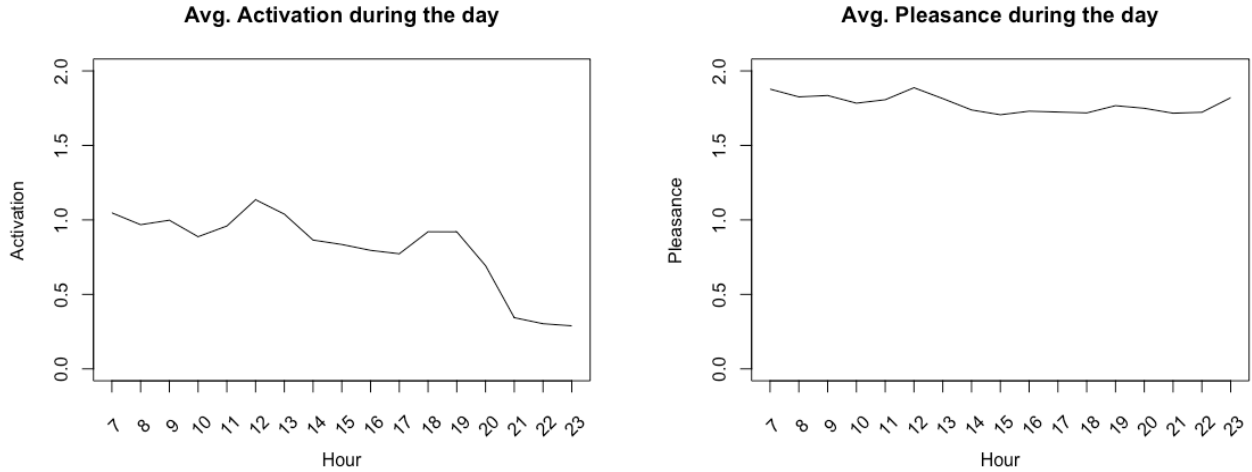

*Figure 6.* Average activation and pleasance during the day

Table 2 indicates that the average pleasance during the experiment was almost twice as high as the average activation. On the other hand the standard deviation of pleasance was lower than the standard deviation of activation. Figure 6 shows that during the day pleasance stays more consistent than activation. This indicates that pleasance is a more consistent state then activation. The later the day gets, the less people tend to be active. Furthermore, it is seen that activation has local peaks at lunch time (11am - 1pm) and work day end time (5pm - 7pm).

---

[2] We have corrected the skeweness in the most recent version of the Happimeter by recalibrating the pleasance scale to (0) unpleasant/sad, (1) pleasant/satisfied, and (2) super pleasant/superhappy. The recalibration is based on the assumption that people are pleasant in their default state (instead of neutral).

| | Pleasance | Activation | Age | Gender Male | Neuroticism | Extraversion | Openness to Experience | Agreeableness | Conscientiousness | Heartrate | Light Level | Activity | VMC | VMC last 4 hours | Temperature | Humidity | Pressure | Wind | Clouds | Weekend/Holiday |
|---|---|---|---|---|---|---|---|---|---|---|---|---|---|---|---|---|---|---|---|---|
| Pleasance | 1*** | | | | | | | | | | | | | | | | | | | |
| Activation | -0.01 | 1*** | | | | | | | | | | | | | | | | | | |
| Age | 0.20*** | 0.07*** | 1*** | | | | | | | | | | | | | | | | | |
| Gender Male | 0.14*** | -0.01 | 0.61*** | 1*** | | | | | | | | | | | | | | | | |
| Neuroticism | 0.08*** | -0.05*** | 0.51*** | -0.45*** | 1*** | | | | | | | | | | | | | | | |
| Extraversion | 0.09*** | 0.08*** | 0.41*** | 0.21*** | -0.28*** | 1*** | | | | | | | | | | | | | | |
| Openness to Experience | 0.02 | -0.05*** | 0.58*** | -0.14*** | 0.56*** | 0.20*** | 1*** | | | | | | | | | | | | | |
| Agreeableness | -0.08*** | 0.00 | 0.38*** | -0.62*** | 0.13*** | 0.17*** | 0.08*** | 1*** | | | | | | | | | | | | |
| Conscientiousness | -0.06*** | 0.05*** | 0.27*** | 0.25*** | 0.09*** | 0.32*** | 0.74*** | -0.28*** | 1*** | | | | | | | | | | | |
| Heartrate | 0.05*** | 0.04** | 0.09*** | 0.13*** | -0.04** | 0.00 | -0.04*** | -0.13*** | 0.00 | 1*** | | | | | | | | | | |
| Light Level | -0.03* | 0.08*** | 0.03** | -0.15*** | -0.11*** | 0.08*** | -0.06*** | 0.25*** | -0.05*** | 0.07** | 1*** | | | | | | | | | |
| Activity | 0.09*** | 0.01 | 0.05*** | -0.32*** | -0.11*** | -0.14*** | -0.41*** | 0.32*** | -0.60*** | 0.18** | 0.17*** | 1*** | | | | | | | | |
| VMC | 0.07*** | 0.11*** | 0.12*** | 0.08*** | 0.02 | -0.14*** | 0.03** | -0.05*** | 0.10*** | 0.29*** | 0.03* | 0.01 | 1*** | | | | | | | |
| VMC last 24 hours | 0.13*** | 0.13*** | 0.17*** | -0.05*** | 0.10*** | 0.03** | 0.18*** | -0.01 | 0.22*** | 0.12*** | 0.06*** | -0.09*** | 0.18*** | 1*** | | | | | | |
| Temperature | 0.03** | 0.08*** | 0.05*** | 0.09*** | -0.22*** | 0.30*** | -0.02 | 0.04*** | 0.12*** | 0.05*** | 0.18*** | -0.03* | -0.04*** | 0.14*** | 1*** | | | | | |
| Humidity | 0.00 | 0.07*** | 0.03** | -0.12*** | 0.22*** | -0.22*** | 0.05*** | -0.07*** | -0.11*** | -0.06*** | -0.19*** | 0.06*** | -0.02 | -0.14*** | -0.77*** | 1*** | | | | |
| Pressure | 0.07*** | 0.08*** | 0.02 | -0.01 | -0.10*** | 0.09*** | -0.01 | 0.24*** | -0.02 | -0.01 | 0.10*** | 0.04*** | 0.10*** | -0.09*** | 0.27*** | 0.26*** | 1*** | | | |
| Wind | 0.02* | 0.08*** | 0.18*** | 0.23*** | 0.06*** | -0.11*** | -0.03* | -0.24*** | 0.03* | 0.07*** | -0.06*** | -0.05*** | 0.05*** | 0.01 | -0.13*** | 0.04*** | -0.42*** | 1*** | | |
| Clouds | 0.13*** | -0.04** | 0.10*** | -0.21*** | 0.25*** | -0.24*** | 0.06*** | 0.02 | -0.12*** | -0.02* | -0.06*** | 0.13*** | 0.02 | 0.08*** | 0.54*** | 0.54*** | 0.27*** | 0.19*** | 1*** | |
| Weekend/Holiday | 0.06*** | -0.07*** | -0.05*** | -0.08*** | 0.03* | -0.12*** | -0.15*** | 0.06*** | -0.23*** | 0.02 | 0.23*** | 0.21*** | -0.02* | -0.01 | 0.08*** | -0.09*** | 0.08*** | -0.12*** | 0.08*** | 1*** |
| Hour | 0.02* | -0.24*** | -0.01 | -0.01 | 0.03** | 0.04** | -0.02 | 0.04** | -0.02 | 0.01 | -0.03** | 0.06*** | -0.01 | 0.30*** | 0.05*** | -0.10*** | -0.05*** | 0.01 | 0.13*** | 0.03* |

*Table 3. Pearson's Correlation Coefficients. (N= 7,861, p\*\*\*<.001, p\*\*<.01, p\*<0.05)*

Table 3 shows that the only correlation above 0.20 is between pleasance and age, which implies a positive linear relationship between age and pleasance. However, the correlations between pleasance and activation with the other variables are on a high significance level (p <.001) for at least one dependent variable. Hence, all parameters are included in our machine-learning model.

Using the software Weka (Holmes, Donkin, & Witten, 1994), we found that the classification made by means of the random forest algorithm (Liaw & Wiener, 2002) produced the best results. Random forests – as part of ensemble learning methods – have the advantage of reducing the problem of overfitting that can arise when using decision trees (Breiman, 2001). In addition, we increased the minimum size of instances per leaf to 50 and validated our model by 10 times cross validation to ensure the absence of overfitting. We trained individual models for each participant and an independent general model for predicting the current mood state. For the individual models, we achieved an accuracy of 94.06% (Cohen's Kappa = 0.75) for pleasance, of 91.59% (Cohen's Kappa = 0.80) for activation, and of 86.68% (Cohen's Kappa = 0.72) for the mood state. The general model achieved an accuracy of 92.04% (Cohen's Kappa = 0.72) for pleasance, of 89.80% (Cohen's Kappa = 0.81) for activation, and of 85.10% (Cohen's Kappa = 0.76) for the Mood State (c.f. Table 4).

|  | Pleasance | | Activation | | Mood State | |
|---|---|---|---|---|---|---|
|  | Accuracy | Cohen's Kappa | Accuracy | Cohen's Kappa | Accuracy | Cohen's Kappa |
| **Individual Model** | 94.06% | 0.75 | 91.59% | 0.80 | 86.68% | 0.72 |
| **General Model** | 92.04% | 0.72 | 89.80% | 0.81 | 85.10% | 0.76 |

*Table 4. Results of 10-fold cross-validation for Random Forest classifications.*

Further, we calculated the feature importances for the general model based on the average impurity decrease and the number of nodes using that attribute. Figures 7 and 8 show that the external influences (including weather) have the largest average impurity decrease even though their number of nodes using them as decision variables is not as high as for other variables. This indicates that the external influences such as the weather set a base mood, which is then further changed by individual context such as the location, the person somebody is with, etc. and indicated by body measures such as heart rate or movement. These findings follow existing research that also identified weather as a predictor for mood (Keller et al. 2005).

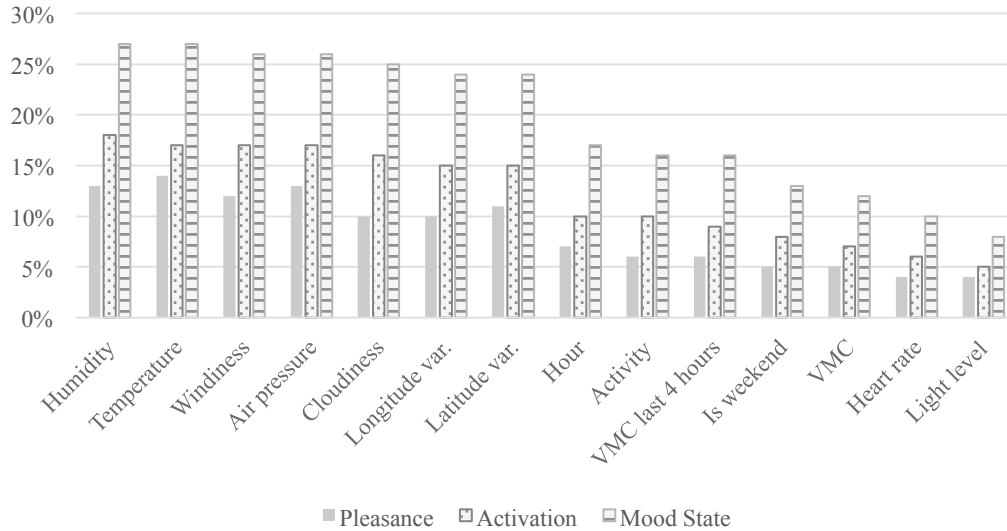

*Figure 7. Average impurity decrease for the most important features of the general model*

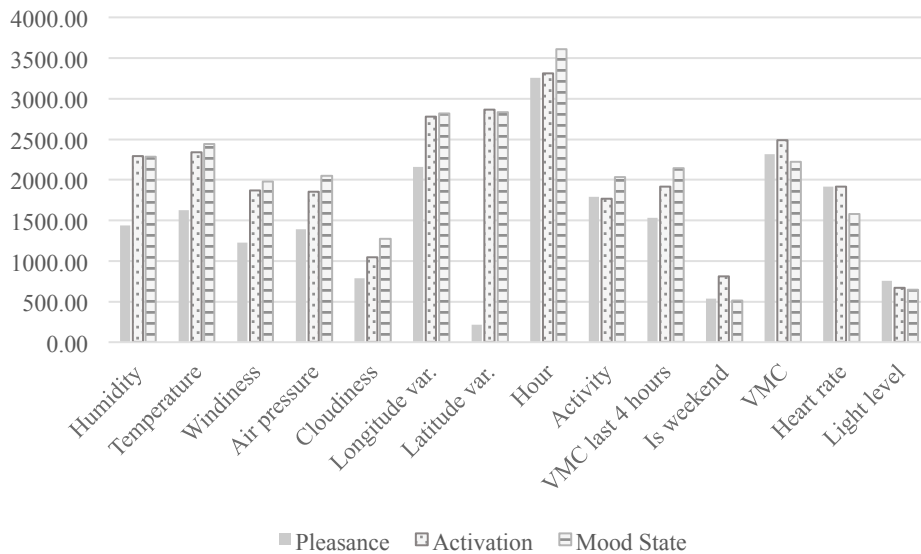

*Figure 8. Number of nodes using the most important features as decision variables in the general model*

## 5. Discussion

Our results support the validity of the measurements collected through smartwatches, used to explore changes in pleasance and activation. We built a general applicable model that can tell the user's mood state with an accuracy of 85.10% and a Cohen's Kappa of 0.76. We also found that there are no strong linear relationships between pleasance/activation and the independent variables. Using average impurity decreases, we distinguished which variables have a stronger impact on pleasance and activation. We found that weather and movement between locations are highly predictive, whereas body measures such as heartrate have lower predictive power. Especially weather – as part of the external influences – is assumed to set a base mood that is further predicted by body measures. Figure 9 summarizes our results, with "+" representing weak impacts and "+++" strong impact on the dependent variables.

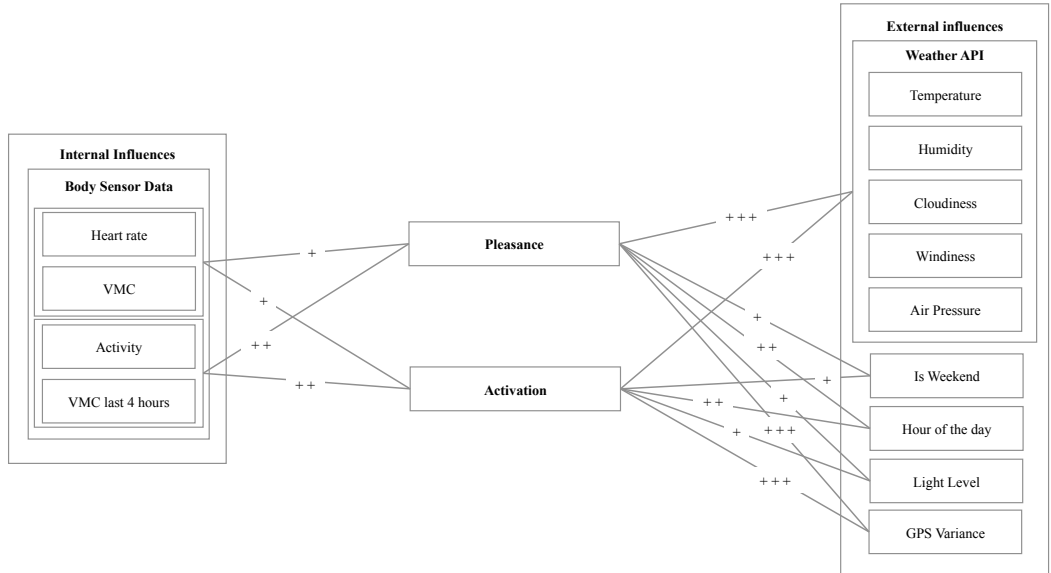

*Figure 9. Significant predictors of Happiness and Activation (based on figures 7 and 8))*

Furthermore, we discovered that pleasance is more consistent than activation, due to its lower standard deviation from its mean value ($SD_{pleasance}$ = 0.4889 vs. $SD_{activation}$ = 0.6384). Activation was generally reported lower ($M_{activation}$ = 0.8989), then pleasance ($M_{pleasance}$ = 1.7688), and decreases during the day. This indicates, that activation can be changed more easily than pleasance.

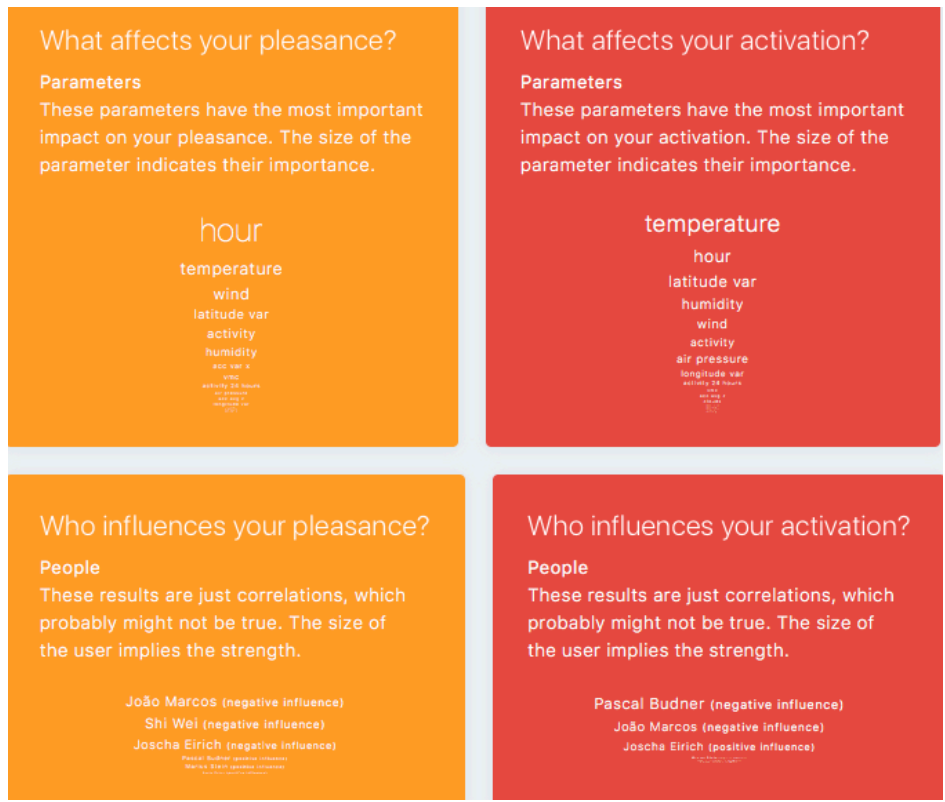

*Figure 10. Insights page of Happimeter Web site*

In addition we also built a prototype system to identify the most predictive variables for each individual person. Figure 10 illustrates the variables that are the strongest predictors for one of the authors. These values are dynamically adjusted depending on the activities and the time of the day. The bottom part of figure 10 also shows who of the friends of an individual has the strongest positive or negative influence on the person, by calculating the size of the correlation between the mood of a person with the friends being present in the same room at the same time.

## 6. Limitations and Future Research

One limitation of this research is the small sample size (60 participants that made frequent mood inputs and 33 that filled out the personality and personal data questionnaires). Also, there might be selection bias in that the participants were people interested in happiness research. Therefore, we suggest replicating our experiment on larger randomly selected samples, including people of different ages and coming from more heterogeneous groups, controlling for other variables such as job, history of mental illness and marital status.

Nevertheless, we think that we have introduced a novel system for tracking and increasing individual happiness, measuring success in the sense of Bill Gates, and tracking Aristotle's goal of increasing happiness by surrounding us with "good friends".

## 7. Acknowledgements

This work has been supported by Philips Lighting as part of an overall MIT "Grand Challenges in Lighting" research grant.